\documentstyle[epsf,12pt]{article}
\begin{document}

\title{Dynamics of Cosmic Strings in Schwarzschild Spacetime}
\author{
Jean-Pierre De Villiers \cr
{\em Theoretical Physics Institute}\cr 
{\em Department of Physics} \cr
{\em University of Alberta} \cr
{\em Edmonton, Canada T6G 2J1}}
\noindent
\maketitle

\vspace {10 pt}
\noindent {\bf Abstract}

Cosmic strings are topological defects thought to have formed early in the life
of the universe. If such objects exist, a study of their interaction with black
holes is of interest.  The equations  of motion of a cosmic string have
the form of highly non-linear wave equations. General analytic solutions,
except for motion in certain backgrounds such as flat and shockwave spacetimes,
remain unknown; consequently, much of the work must be carried out numerically.
To do this, an implicit finite difference scheme was developed that involves
solving large block tridiagonal systems.  This paper discusses the numerical
method, its validation against analytic and semi-analytic solutions, and
preliminary results on the interaction of a cosmic string with Schwarzschild
black holes.

\section{Introduction}

This paper
\footnote{Preprint: Alberta Thy 12-97. To appear in
Proceedings of Seventh Canadian Conference on General Relativity and
Relativistic Astrophysics (Fall 1997)} 
discusses numerical solutions to the equations of motion
of a cosmic string in black hole spacetimes. 
Strings can be either closed (a loop) or open. Open strings can be of infinite
length, or they can be terminated by monopoles; strings can also terminate on
the horizon of a black hole (due to earlier capture). From the numerical point
of view, cosmic strings of finite length are preferable since they lead
directly to finite computational domains. The goal is to model the string as a
segment terminated by massive end points, taking the limit where the masses go
to infinity to simulate an infinite open string. 

A cosmic string is an extended object under tension; motion of the string near
a black hole represents the resultant of the competing influences of tension
and gravity. The results presented here concern the deformations induced by the
gravitational field of Schwarzschild black holes. These results will
be compared to the behaviour of dust strings, one-dimensional distributions of
test particles that mimic a string with no tension.
\section{Equations of Motion of a Cosmic String}

The equations of motion for a cosmic string with massive end points are derived
from an action functional.  Solutions to these equations describe the world 
surface of the string as functions ${x}^{\mu}\,({\xi}^{0},{\xi}^{1})$ of the 
string parameters ${\xi}^{A}  (A=0,1)$ conventionally shown as
$({\xi}^{0},{\xi}^{1})\equiv(\tau,\sigma)$.  The derivation builds upon two sources,
Larsen and Frolov \cite{Larsen94}, who obtain equations of motion for open
strings in arbitrary backgrounds, and Barbashov \cite{Barbashov77}, who obtains 
equations of motion for strings with massive end
points in Minkowski spacetime.  The action for a string terminated by massive
end points in an arbitrary spacetime is taken to be,
\begin{eqnarray}\label{m.5}
S\left[{x}^{\mu},{h}_{AB},{\sigma}_{i}\right]& = & -\mu \int_{{\tau}_{1}}^{{\tau}_{2}}
{d\tau \int_{{\sigma}_{1}(\tau)}^{{\sigma}_{2}(\tau)}
{d\sigma\,\sqrt{-h} \,h^{AB}\,G_{AB}}}\\
\nonumber & - &
\sum_{i=1}^{2}{{{m}_{i} \over 2}\int_{{\lambda}_{1}}^{{\lambda}_{2}}
{d\lambda\,\,{g}_{\mu \nu}\,
{d{x}^{\mu} \over d\lambda}\, 
{d{x}^{\nu} \over d\lambda}}}
\end{eqnarray}
where $\mu$ is the string tension; ${\sigma}_{i}(\tau)$ represent the
motion of the end points in the 2D parameter space of the string;
${h}_{AB}$ is the metric of the string parameter space, with
determinant $h$; $G_{AB} \equiv g_{\mu \nu}\,{\partial {x^{\mu}}\over 
\partial {\xi}^{A}}\,{\partial {x^{\nu}}\over \partial {\xi}^{B}}$ 
is the induced metric on the string worldsheet;
and $\lambda \equiv \lambda(\tau)$ is the most general parametrization
for the end points whose motion is described by 
${x}^{\mu}\left(\lambda(\tau),{\sigma}_{i}(\tau)\right)$.

The end points are taken to represent the mass of a longer
(potentially infinite) string lying outside the region of interest; the motion
of these points represents a boundary condition for the section of the string
worldsheet under study. In the limit of infinite mass, these end points move
along geodesics.  

The equations of motion are obtained by taking the $h_{AB}$, $x^{\mu}$, and
${\sigma}_{i}$ variations of the action. The details of this derivation
are given elsewhere \cite{DeVilliers}. The internal string coordinates are 
chosen such
that $\tau$ is a time-like coordinate, and $\sigma$ a spatial coordinate
used to parametrize points along the string. The resulting equations of motion
for the interior of the string and boundary conditions are,
\begin{equation}\label{m.15}
{{\partial}^{2}\,{x}^{\mu} \over \partial\,{\tau}^{2}}+
{\Gamma}^{\mu}_{\rho \eta}\,
{\partial\,{x}^{\rho} \over \partial\,\tau}\,
{\partial\,{x}^{\eta} \over \partial\,\tau} =
\cases{
{{\partial}^{2}\,{x}^{\mu} \over \partial\,{\sigma}^{2}}+
{\Gamma}^{\mu}_{\rho \eta}\,
{\partial\,{x}^{\rho} \over \partial\,\sigma}\,
{\partial\,{x}^{\eta} \over \partial\,\sigma}
\,;\,(-{\pi \over 2} < \sigma < {\pi \over 2})\\
\cr
\cr
0\,;\,(\sigma = \pm {\pi \over 2})
}
\end{equation}
along with constraints, 
\begin{eqnarray}\label{d.24}
{g}_{\mu \nu}\,
\left[{\partial\,{x}^{\mu} \over \partial\,\tau}\,
      {\partial\,{x}^{\nu} \over \partial\,\tau}+
      {\partial\,{x}^{\mu} \over \partial\,\sigma}\,
      {\partial\,{x}^{\nu} \over \partial\,\sigma}
\right]
 & = & 0\,;\,(-{\pi \over 2} < \sigma < {\pi \over 2})\\
\nonumber {g}_{\mu \nu}\,
\left[{\partial\,{x}^{\mu} \over \partial\,\tau}\,
      {\partial\,{x}^{\nu} \over \partial\,\sigma}
\right]
 & = & 0\,;\,(-{\pi \over 2} < \sigma < {\pi \over 2})\\
\nonumber {d \over d\tau}\left({g}_{\mu \nu}\,
\left[{d\,{x}^{\mu} \over d\,\tau}\,
      {d\,{x}^{\nu} \over d\,\tau}
\right]\right)
 & = & 0\,;\,( \sigma = \pm {\pi \over 2})
\end{eqnarray}
which are used as checks on the quality of the numerical solutions to
(\ref{m.15}).

\section{Discretization of the Equations of Motion}

The above equations have the form of non-linear wave equations. Ames
\cite{Ames65} describes a discretization due to Von Neumann for the linear wave
equation, ${u}_{\tau \tau} = {u}_{\sigma \sigma}$, which uses a standard
second-order centered difference formula for evaluating the time derivative at
the discrete grid point ${u}\left({\sigma}_{i},{\tau}_{j}\right)$, and a
weighted average (weighting parameter $\lambda$) of centered spatial
differences at three adjacent time steps, ${\tau}_{j-1}$, ${\tau}_{j}$, and
${\tau}_{j+1}$. This averaging gives rise to a 9-point implicit scheme. 
Von Neumann's discretization must be extended to handle a system of
equations and the non-linear terms containing Christoffel symbols. To 
simplify the notation, denote 
${x}^{\mu}\left({\sigma}_{i},{\tau}_{j}\right) \equiv {\bf x}_{i,j}$.
The second-order time derivative is discretized using a centered difference
formula on a non-uniform mesh (to accommodate the iterative scheme
described below),
\begin{equation}\label{d.9c}
{\left({{\partial}^{2}\,{\bf x} \over \partial\,{\tau }^{2}}\right)}_{i,j} \approx 
{2 \over \Delta{\tau}_{j} + \Delta{\tau}_{j+1}}\,
\left\{ 
{1  \over  \Delta{\tau}_{j+1}}\,\left( {\bf x}_{i,j+1} - {\bf x}_{i,j}  \right) -
{1 \over  \Delta{\tau}_{j}}\,\left( {\bf x}_{i,j} - {\bf x}_{i,j-1}  \right)
\right\}
\end{equation}
while the second-order spatial derivative is discretized using the
standard centered difference, denoted ${D}^{2}{\bf x}_{i,j}$,
as an average of three time levels, 
\begin{equation}\label{d.8}
{\left({{\partial}^{2}\,{\bf x} \over \partial\,{\sigma}^{2}}\right)}_{i,j} 
\approx 
{\lambda \over {\left( \Delta\,{\sigma} \right)}^{2} }\,{D}^{2}{\bf x}_{i,j+1}  + 
{\left( 1-2\,\lambda \right) \over {\left( \Delta\,{\sigma} \right)}^{2} }\,
{D}^{2}{\bf x}_{i,j} + 
{\lambda \over {\left( \Delta\,{\sigma} \right)}^{2} }\,{D}^{2}{\bf x}_{i,j-1}
\end{equation}

The non-linear spatial term, denoted $H$ for compactness, is also discretized
using the averaging method, ${\left(H\right)}^{n}_{i,j} \approx \lambda \,{H}^{n}_{i,j+1} + 
\left(1-2\,\lambda \right)\,{H}_{i,j}+\lambda\,{H}_{i,j-1}$, where the
upper index is an iteration index.  The contributions at the $j$ and $j-1$
level can be obtained directly from centered differences, evaluating the
Christoffel symbols (using analytic expressions) at ${\bf x}_{i,j}$ and ${\bf
x}_{i,j-1}$. The $j+1$ term is linearized through a Taylor expansion,
${H}^{n+1} _{i,j+1} \cong  {H}^{n} _{i,j+1} +  {J}({{\bf x}}^{n}_{i,j+1})
\bullet\, \Delta{{\bf x}}^{n}_{i,j+1}$ where $ \Delta{{\bf x}}^{n}_{i,j+1}$
represents the difference between the iterated solutions at the $j+1$
time-level,  $\Delta{{\bf x}}^{n}_{i,j+1} = {{\bf x}}^{n+1}_{i,j+1} - {{\bf
x}}^{n}_{i,j+1}$, and ${J}\left({{\bf x}}\right)$ is the Jacobian of the
n-th iteration of ${H}^{n}_{i,j+1}$. The temporal non-linear term is expanded
using centered differences on a non-uniform temporal mesh and rearranged using
the $\Delta$-notation introduced above. This term is linearized directly by
discarding terms of order ${\Delta{\bf x}}^{2}$.

The various difference expressions are combined and
the terms involving the iterated solution, $\Delta\,{\bf x}^{n}_{*,j+1}$,
isolated on the right hand side. The system reduces to an equation
for the components of a block-tridiagonal system,
\begin{equation}\label{d.14}
{A }^{n}_{i-1,j+1}\bullet \Delta\,{\bf x}^{n}_{i-1,j+1} + 
{B }^{n}_{i,j+1} \bullet \Delta\,{\bf x}^{n}_{i,j+1} +
{C }^{n}_{i+1,j+1} \bullet \Delta\,{\bf x}^{n}_{i+1,j+1} = {d }^{n}_{i,j+1}
\end{equation}
where ${A}$, ${B}$, and ${C}$ are $4 \times 4$ matrices and ${d}$ is a
4-vector.  The block-tridiagonal system expresses an implicit relationship
between the iterated solution at all points on the spatial grid (of N points)
and reduces to the matrix expression ${T }^{n} \bullet \Delta\,\vec{{\bf
x}}^{n} = \vec{d}^{n}$  where $T$ in an $N \times N$ matrix whose elements are
$4 \times 4$ matrices, and $\vec{{\bf x}}^{n}$ and $ \vec{d}^{n}$ are vectors
of length N whose elements are 4-vectors. The iterated solution for the new
time level is found by inverting matrix T; the process is repeated until the 
solution converges (i.e. produces a zero d-vector). Failure to converge
triggers a time-step reduction mechanism; if this mechanism fails to find a
suitable new step size, the solution is stopped. The constraint equations
(\ref{d.24}) are invoked periodically to assess the quality of the solution by
computing an average value (and standard deviation) for the constraints over
the spatial grid. 
\section{Testing and Using the Solver}

The finite difference expressions were coded and tested against an analytic
flat space solution (in spherical-polar coordinates so as to test the
non-linear portions of the code) and a semi-analytic shockwave solution. The
former test established the accuracy of the method, the latter tested the
iterative scheme at the ultra-relativistic extreme. 

\subsection{Minkowski Spacetime}

An analytic expression is easily obtained for a cosmic string of length L moving
at constant velocity $v$ in Minkowski spacetime. In Cartesian coordinates, 
the string is oriented along the z-axis, such
that $Z(\tau,\sigma) = L\,\sigma/\pi$,   ($-{\pi \over 2} \le \sigma \le {\pi
\over 2}$), with motion taking place along the
x-axis. The motion is described by,
\begin{equation}\label{d.23b}
{\bf x}(\tau,\sigma) = 
\left({L \over \pi}\,\cosh{(\beta)}\,\tau+{T}_{0},
{L \over \pi}\,\sinh{(\beta)}\,\tau+{X}_{0},{Y}_{0},{L \over \pi}\,\sigma \right)
\end{equation}
where ${X}_{0}$, ${Y}_{0}$, and ${T}_{0}$ represent arbitrary initial values.
The parameter $\beta$ represents the rapidity of a string moving with velocity
$v = \tanh{\beta}$. This solution was coded in spherical-polar coordinates to
exercise the non-linear portions of the code, and is compared to the numeric
solution in Figure 1, where the absolute relative error averaged over
the spatial grid is plotted as a function of string proper time, $\tau$. The
graph covers roughly 10,000 time steps, and the slope confirms that the 
method is second-order in the time step size $\Delta\tau$.

This solution (expressed in Schwarzschild coordinates, $(t,r,\theta,\phi)$) is
also used as initial data for all numerical work, since black hole spacetimes
of interest are asymptotically flat.  The initial time ${t}_{0}$ is taken to be
zero. The value of ${r}_{0}$ sets the initial distance from the black hole and
is chosen to make the error of the analytic solution small. The value of 
${\phi}_{0}$ sets the initial impact parameter $b$ ($b = {r}_{0}
\sin{{\phi}_{0}}$) relative to the black hole. The string is oriented
symmetrically about the equatorial plane.  

\subsection{Shockwave Spacetime}

The shockwave spacetime arises when an extremely relativistic string
passes near a black hole. In this situation, the string effectively
propagates in Minkowski spacetime up to the point where it encounters
the sharply defined gravitational potential of the black hole. 

In the limit where string velocity $v \rightarrow c$, the Schwarzschild 
and Kerr metrics can be reduced to a shockwave form (Hayashi and Samura, 
\cite{Hayashi94}), 
\begin{equation}\label{sw.1} 
d\,{s}^{2} = -
du\,dv + d{y}^{2} + d{z}^{2} -  4\,p\,g(\rho)\,\delta(u)\,d{u}^{2}
\end{equation}
where,  
\begin{equation}\label{sw.2} 
u = t - x; 
v = t + x; 
p =\gamma\,{M}_{BH};
\rho = \sqrt{{y}^{2}+{z}^{2}} 
\end{equation} 
and, for the Schwarzschild case,
\begin{equation}\label{sw.3} 
g(\rho) = \ln{{\rho}^{2}}-{\rho \over 4\,{M}_{BH}}
\left[\pi-4\, \sqrt{1 - {4\,{M}_{BH}^{2} \over {\rho}^{2}}}\,
\arctan{\sqrt{\left({\rho + 2\,{M}_{BH} \over \rho - 2\,{M}_{BH}}\right)}} \right]
\end{equation} 

In this metric, the string is propagating in the x-direction and
the coordinates y and z are transverse to the motion.
Spacetime is everywhere flat, except at $u=0$, so the
flat-space constant velocity solutions are used as a starting point. 
The equations of motion of strings in a shockwave background can be
expressed as a Fourier series (adapted from Amati and Klimcik, \cite{Amati88}),
\begin{eqnarray}\label{sw.3a} 
\nonumber U(\sigma,\tau) & = & {p}^{u}\,\tau\\
\nonumber V(\sigma,\tau) & = & {v}_{0} + {p}^{v}\,\tau +
i\,\sum_{n \ne 0}{{{\alpha}^{v}_{n} \over n}\,{e}^{-i n \tau}\cos(n \sigma)}\\
\nonumber Y(\sigma,\tau) & = & b + {p}^{y}\,\tau +
i\,\sum_{n \ne 0}{{{\alpha}^{y}_{n} \over n}\,{e}^{-i n \tau}\cos(n \sigma)}\\
Z(\sigma,\tau) & = & {L \over \pi} \left(\sigma - {\pi \over 2} \right) + 
{p}^{z}\,\tau +
i\,\sum_{n \ne 0}{{{\alpha}^{z}_{n} \over n}\,{e}^{-i n \tau}\cos(n \sigma)}
\end{eqnarray} 
where ${p}^{\alpha}$ is the $\alpha$-component of the string's 4-momentum,
and the Fourier coefficients for the $Y$ and $Z$ solutions are given by,
\begin{eqnarray}\label{sw.3b} 
{\alpha}^{i}_{n} & = & {{p}^{u} \over 2 \pi}
{\int}^{\pi}_{0}{{\partial}_{i}f(X(\sigma,0))\cos(n \sigma)\,d\sigma}
\end{eqnarray} 
where $f(X(\sigma,0))=-4\,p\,g(X(\sigma,0))$, and ${\alpha}^{i}_{n}$ is
evaluated numerically.

It is possible to compare this series expansion with numerically-generated
solutions for a string moving at ultra-relativistic velocity. The most
interesting comparison is made for projections of the worldsheet onto the Y-Z
plane.  Figure 2 shows the numerically generated solution, the shockwave
series solution (to 300 terms), and a numerically generated dust string
solution where an array of test particles with identical initial data as
the cosmic string are propagated using the geodesic equation for
test particles.  

\subsection{Close Encounters}

A cosmic string straying too close to a black hole will become trapped; the
conditions under which this happens, expressed in terms of the 2D parameter
space of initial velocity $v$ and impact parameter $b$, can be used to generate
a plot of the critical impact parameter (a line in b-v space). These results
are discussed in a separate paper \cite{DeVilliers}. The previous section
discussed the excitation of ultra-relativistic strings; here we consider the
case of a string moving more slowly. The study of geodesics of particles shows
that for impact parameters near the critical value, particles can execute a
partial orbit around the black hole before escaping. Figure 3 shows the
trajectory of a slow string with near-critical impact parameter, displaying a
time-sequence for the portion of a string in the immediate vicinity of the
equatorial plane, which shows the string attempting a partial orbit before
being pulled back into its original direction of motion. 
\section{Discussion}

In this paper, equations of motion for a cosmic string were derived from an
action that deals with a finite string terminated by massive
end points. The end points are taken to represent the mass of segments
(potentially infinite) lying outside the region of interest; the motion
of these points represents a boundary condition for the section of the string
worldsheet under study. In the limit of infinite mass, these end points move
along geodesics.  The equations of motion for the interior of the worldsheet
have the form of highly non-linear wave equations.

A numerical method was developed based on
Von Neumann's discretization of the wave equation. The discretized equations
are linearized and give rise to a block-tridiagonal system of equations that
must be solved iteratively to advance the solution in time. 

The validation of this method against analytic solutions in Minkowski
spacetime and semi-analytic solutions in shockwave spacetimes was discussed, as
well as a comparison against dust strings. The excitation of a slowly moving
string with impact parameter near the critical parameter for trapping was
also shown.

These results illustrate the role of tension in determining the dynamics of the
string. At ultra-relativistic speeds, a string develops a strong excitation in
the direction normal to its original motion. The cosmic string and the dust
string both exhibit loop formation immediately following the encounter, as
portions of the strings nearest the equatorial plane cross over to the other
side. Whereas the portions of the dust string undergoing this cross-over
continue in their new direction, those portions of the cosmic string that have
undergone a cross-over eventually recross the equatorial plane, giving the
worldsheet a kink-like profile that propagates outward along the string. At
lower velocities, with near-critical impact parameters, loop formation is also
observed, this time in the direction of motion. The cosmic
string performs a partial orbit, but eventually gets pulled back in the
original direction of motion, giving rise to the looping structure. The time
sequence shown in Figure 3 is for a string with initial impact parameter of
2.28 ${r}_{g}$; this value is less that the minimum impact parameter of $\sim
2.6\,{r}_{g}$ for particles (which occurs at the ultra-relativistic limit).
Numerical studies of the trapping of cosmic strings has shown that they are
much  more difficult to trap than particles; this will be discussed in greater
detail in \cite{DeVilliers}.

\newpage
\begin{figure}
\epsffile{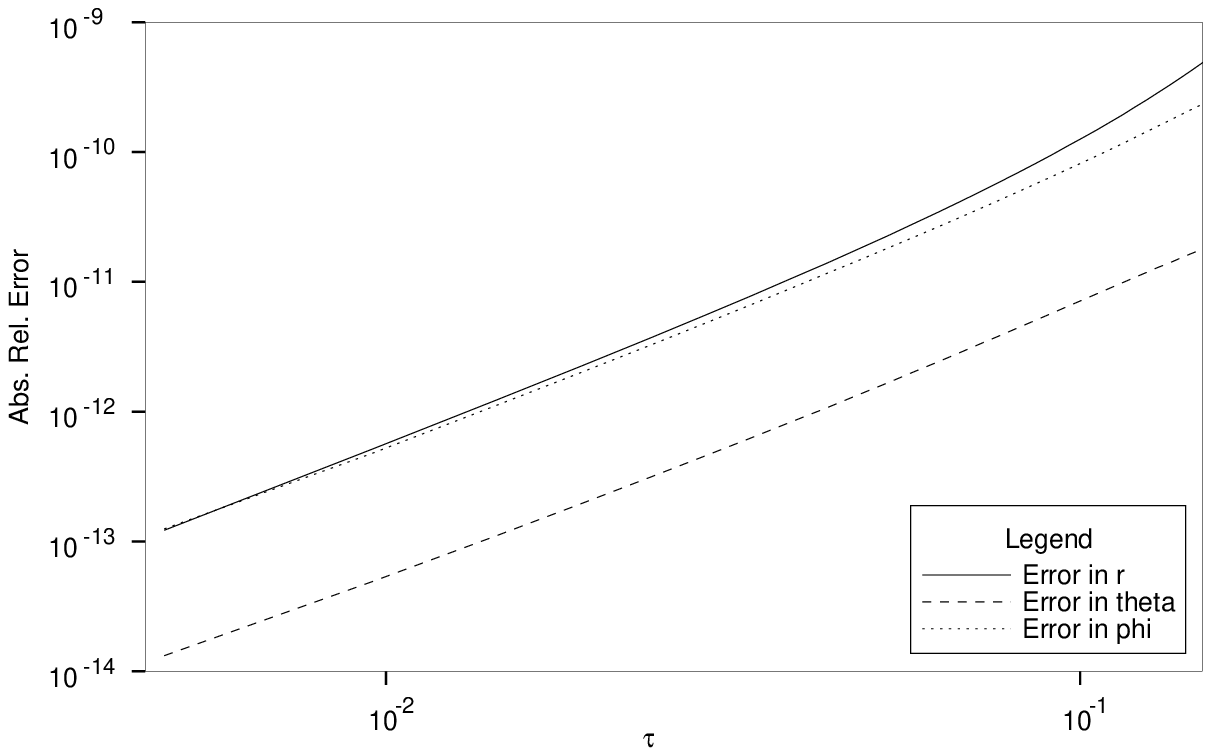}
\caption{Comparison of numerical solution to analytic solution in
Minkowski spacetime.}
\end{figure}

\begin{figure}
\epsffile{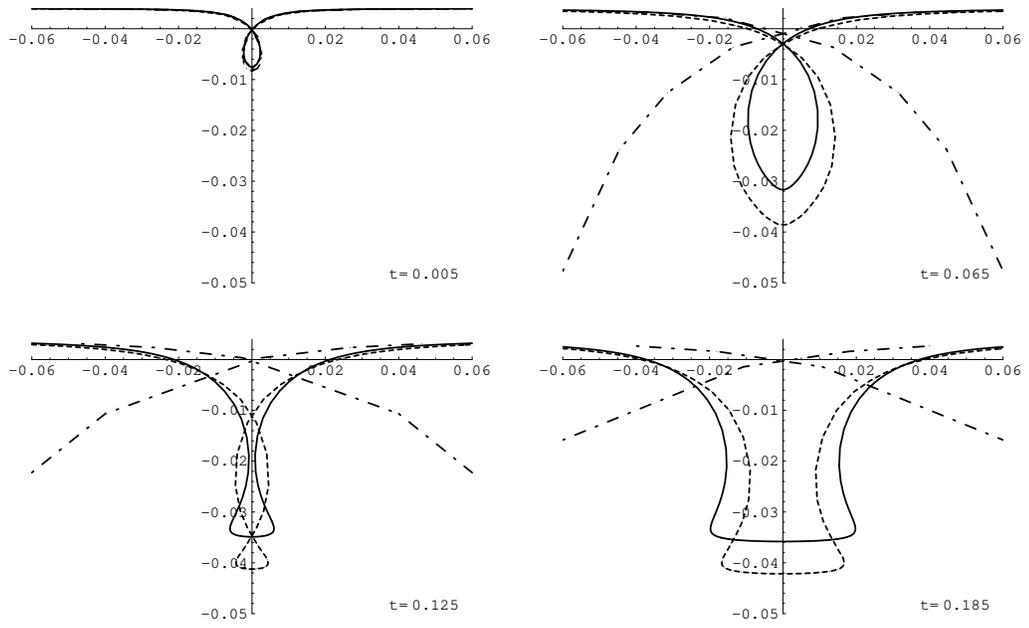}
\caption{Comparison of numerical solution (solid line) to shockwave (dashed)
and dust (dash-dot) solutions.
Initial velocity, 0.995c, impact parameter 4 ${r}_{g}$.}
\end{figure}

\begin{figure}
\epsffile{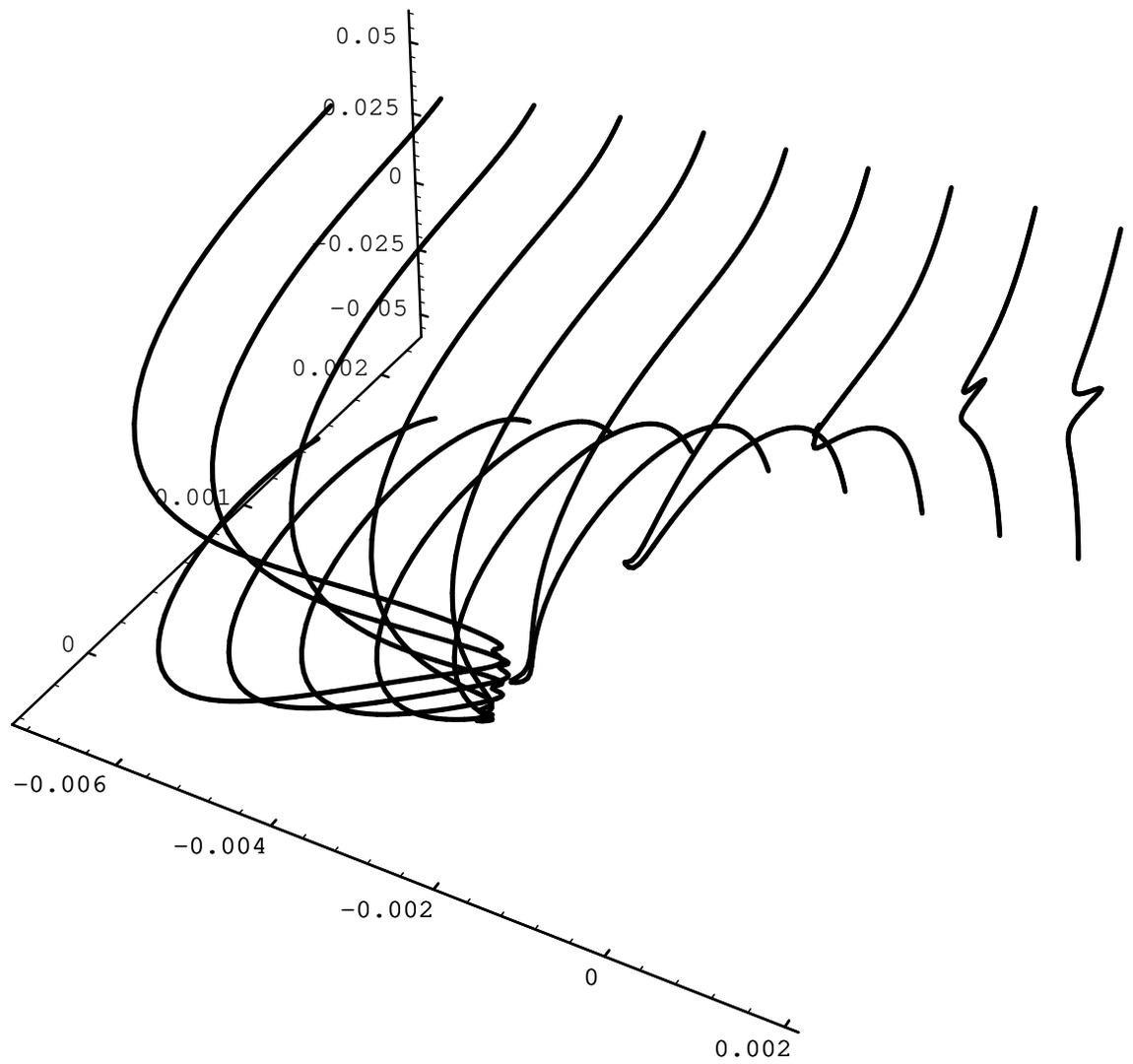}
\caption{Time sequence of string close encounter. Black hole lies at
origin of coodinate system. Initial velocity, 0.125c, 
impact parameter 2.28 ${r}_{g}$.}
\end{figure}

\end{document}